\documentclass[english,aps,floatfix,final,notitlepage,oneside,onecolumn,nobibnotes,nofootinbib,superscriptaddress,showpacs]{revtex4}
\usepackage[T1]{fontenc}
\usepackage[latin9]{inputenc}
\usepackage{graphicx}

\makeatletter

\providecommand{\tabularnewline}{\\}

\makeatletter

\usepackage{epsfig}

\usepackage{amsfonts}

\usepackage{epsfig}

\usepackage{graphics}

\usepackage{babel}

\begin{document}

\title{ Study of exclusive processes $e^{+}e^{-}\to VP$. }

\author{V.V. Braguta}

\email{braguta@mail.ru}

\author{A.K. Likhoded}

\email{Likhoded@ihep.ru}

\author{A.V. Luchinsky}

\email{Alexey.Luchinsky@ihep.ru}

\affiliation{Institute for High Energy Physics, Protvino, Russia}

\begin{abstract}
This paper is devoted to consideration of the hard exclusive processes $e^+e^- \to VP$, where $V=\rho,\phi;~P=\eta,\eta'$. Experimental measurement of the cross section of the process $e^+ e^- \to \phi\eta$ at BaBar collaboration at large center mass energy $\sqrt s=10.6$ GeV  and some low energy experimental data $\sqrt s \sim 2-4$ GeV  give us the possibility to study the cross section in the broad energy region. As the result, we have determined the asymptotic behavior of the cross section of $e^+ e^- \to \phi\eta$ in the limit $s \to \infty$, which is in agreement with perturbative QCD prediction. Assuming that the same asymptotic behavior is valid for the other processes under consideration and using low energy experimental data we have predicted the cross sections of these processes  at energies $\sqrt s=3.67,~10.6$ GeV. In addition, we have calculated the cross sections of these processes at the same energies within perturbative QCD. Our results are in agreement with available experimental data.
\end{abstract}


\maketitle

\section{Introduction}

Exclusive hadron production in high energy electron-positron annihilation is a very interesting task for theoretical and experimental investigations. The presence of high energy scale $\sqrt{s}$ that is much greater than typical hadronic scale allows one to separate the amplitude of such processes into hard part (creation of quarks at very small distances) and soft part (subsequent hadronization of these quarks into experimentally observed mesons at larger distances). The first part of the amplitude can be calculated within perturbative QCD. The second part of the amplitude is described by distribution amplitudes (DA), which contain nonperturbative properties of final hadrons.

The description of hard exclusive hadron production within this pattern gives some very interesting predictions of the properties of hard exclusive processes \cite{Chernyak:1983ej,Lepage:1980fj}. One of such prediction is the asymptotic behavior of the amplitudes and cross sections of hard exclusive processes in the limit $s\to\infty$ \cite{Chernyak:1977fk,Chernyak2,Chernyak3}. It turns out that this behavior is determined by the perturbative part of amplitude and quantum numbers of final hadrons and does not depend on DAs of final hadrons.

To give quantitative prediction for the cross section of hard exclusive process one needs to know DAs. It is interesting to note that if DAs are known the theory can equivalently well predict the cross sections for the production of hadrons composed of light ($u,d,s$ quarks) or heavy quarks ($b,c$ quarks). There is well known example of exclusive process with heavy quarkonia production $e^{+}e^{-}\to J/\Psi\eta_{c}$ measured at Belle \cite{Abe:2002rb} and BaBar \cite{Aubert:2005tj} collaborations. This process was extensively studied in many papers \cite{Chao1,Chao2,Chao3,Chao4,Choi:2007ze,Bondar:2004sv,Braguta:2005kr,Braaten:2002fi,Bodwin:2007ga,Ma:2004qf} within different approaches, what led to a better understanding of charmonia properties and production processes. It is important to note that the approach to hard exclusive processes described above leads to a reasonable agreement with the experiments.

In this paper we study the processes $e^{+}e^{-}\to VP$, where $V=\rho,\phi;~P=\eta,\eta'$.  Experimental measurement of the cross section of the process $e^+ e^- \to \phi\eta$ at BaBar collaboration at large center mass energy $\sqrt s=10.6$ GeV  and some low energy experimental data $\sqrt s \sim 2-4$ GeV  allow us  to study the cross section of this process in the broad energy region. Our first purpose is to use these data in order to determine the asymptotic behavior of the cross section of the process $e^{+}e^{-}\to \phi \eta$. Assuming that the same asymptotic behavior is valid for the other processes under consideration and using low energy experimental data one can predict the cross sections of the $e^{+}e^{-}\to VP$ at energies $\sqrt s=3.67,~10.6$ GeV. In addition, we  apply perturbative QCD approach  to estimate the values of the cross sections for the processes under consideration.

This paper is organized as follows. In the next section we  analyze the experimental data for the process $e^+ e^- \to \phi\eta$ and determine the asymptotic behavior of this cross section. Then we apply the result of this study to predict the cross sections of the other processes under consideration at the center mass energies $\sqrt s=3.67,~10.6$ GeV. In section \ref{sec:Light-Cone} we give theoretical predictions for the cross sections $\sigma(e^{+}e^{-}\to\rho^{0}\eta)$, $\sigma(e^{+}e^{-}\to\rho^{0}\eta')$, $\sigma(e^{+}e^{-}\to\phi\eta)$ and $\sigma(e^{+}e^{-}\to\phi\eta')$ at $\sqrt{s}=3.67$ GeV and 10.6 GeV and compare them with available experimental data. The final section is devoted to the discussion of the results of this paper.

\section{The asymptotic behavior\label{sec:Asymptotic-Behavior}.}

The amplitude of the process involved can be written in the following
form: \begin{eqnarray*}
\mathcal{M}(e^{+}e^{-}\to VP) & = & 4\pi\alpha\frac{\bar{v}(q_{1})\gamma^{\mu}u(q_{2})}{s}\left\langle V(p_{1},\lambda)P(p_{2})\left|J_{\mu}^{em}\right|0\right\rangle ,
\end{eqnarray*}
where $\alpha$ is the electromagnetic coupling constant, $u(q_{2})$ and $\bar{v}(q_{1})$ are electron and positron bispinors, $s=(q_{1}+q_{2})^{2}$ is the invariant mass of $e^{+}e^{-}$ system squared and $J_{\mu}^{em}$ is electromagnetic current. The matrix element $\left\langle VP\left|J_{\mu}^{em}\right|0\right\rangle $ can be parameterized by the only formfactor $F(s)$:
\begin{eqnarray}
\left\langle V(p_{1},\lambda)P(p_{2})\left|J_{\mu}^{em}\right|0\right\rangle  & = & ie_{\mu\nu\alpha\beta}\epsilon_{\lambda}^{\nu}p_{1}^{\alpha}p_{2}^{\beta}F(s),\label{eq:Matr}
\end{eqnarray}
where $\epsilon_{\lambda}^{\nu}$ is the polarization vector of meson
$V$. The cross section of the process under consideration equals
\begin{eqnarray}
\sigma(e^{+}e^{-}\to VP) & = & \frac{\pi\alpha^{2}}{6}\left(\frac{2|\mathbf{p}|}{\sqrt{s}}\right)^{3}\left|F(s)\right|^{2}.\label{eq:SIGMA}
\end{eqnarray}
In the last formula $\mathbf{p}$ is the momentum of the vector meson $V$ in the center mass frame of final mesons.

\begin{figure}
\begin{centering}
\includegraphics{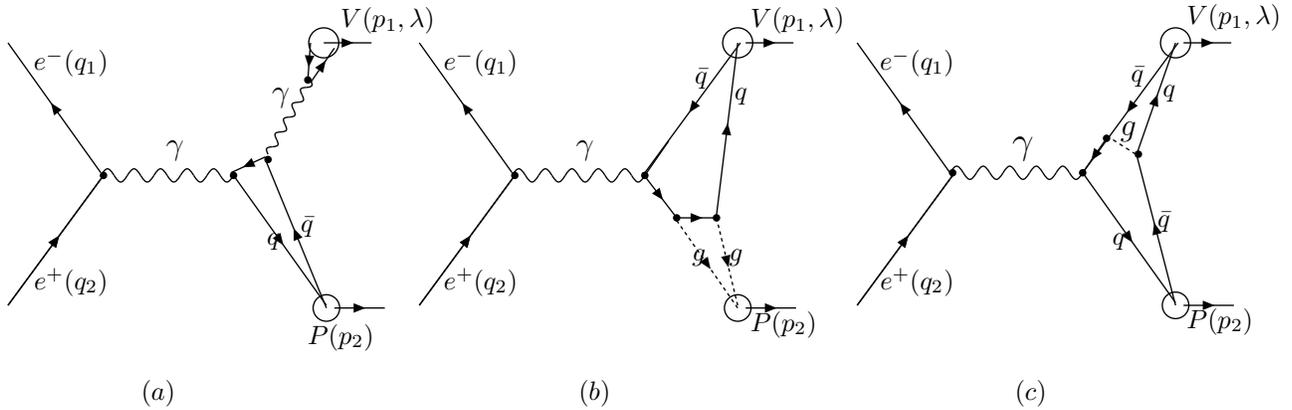}
\par\end{centering}
\caption{Typical diagrams for the $e^{+}e^{-}\to VP$ process.\label{fig:diags}}
\end{figure}

In this section we will be interested in the asymptotic behavior of the formfactor $F(s)$ in the high energy region. Typical diagrams of the process under consideration are shown in Fig. \ref{fig:diags}. The asymptotic behavior of the diagram shown in Fig. \ref{fig:diags}a is $\sigma\sim1/s^{2}$. At extremely large energies this diagram gives the dominant contribution. However, the amplitude of this diagram is suppressed by the smallness of the electromagnetic coupling constant and our study shows that in the energy region analyzed in this paper the contribution of this diagram is negligible. Further let us consider the diagrams shown in  Fig. \ref{fig:diags}b,c. According to perturbative QCD \cite{Chernyak:1983ej} amplitudes (\ref{eq:Matr}) from such diagrams have the following asymptotic behavior
\begin{eqnarray*}
\left\langle H_{1}(p_{1},\lambda_{1})H_{2}(p_{2},\lambda_{2})\left|J_{\mu}^{em}\right|0\right\rangle  & \sim & \left(\frac{1}{\sqrt{s}}\right)^{|\lambda_{1}+\lambda_{2}|+1},
\end{eqnarray*}
where $H_{1}$ and $H_{2}$ are mesons with momenta $p_{1}$, $p_{2}$ and helicities $\lambda_{1}$ and $\lambda_{2}$. For the process under consideration $H_{1}$ and $H_{2}$ are the vector and pseudoscalar mesons respectively. The helicity of the pseudoscalar meson is, obviously, $\lambda_{2}=0$. Because of antisymmetric tensor in (\ref{eq:Matr}), longitudinal polarization of the vector meson is forbidden and it is transversely polarized ($\lambda_{1}=\pm1)$. So, we have $F(s)\sim1/s^{2}$, $\sigma(e^{+}e^{-}\to VP)\sim1/s^{4}$. On the other hand, in papers \cite{Lu:2007hr,Gerard:1997ym}, it was stated that experimental data can be described only by the dependence $\sigma\sim1/s^{3}$.  To clarify this situation we will parameterize the formfactor $F(s)$ by the expression
\begin{eqnarray}
F(s) & = & \frac{a_{n}(s)}{\left(\sqrt{s}\right)^{n}},\label{eq:F}
\label{asympt}
\end{eqnarray}
where $a_{n}(s)$ slightly depends on $s$ due to power and logarithmic corrections to the leading asymptotic behavior. Later in this section we will neglect such dependence. This approximation allows us to fix the constants $a_{n}$ from the low energy data and predict the cross sections at $\sqrt{s}=3.67$ GeV and 10.6 GeV. We check three different hypothesis ($n=3,4$ and 5) and compare the results with existing experimental data.

\begin{figure}[p]
\begin{centering}
\includegraphics[scale=0.7]{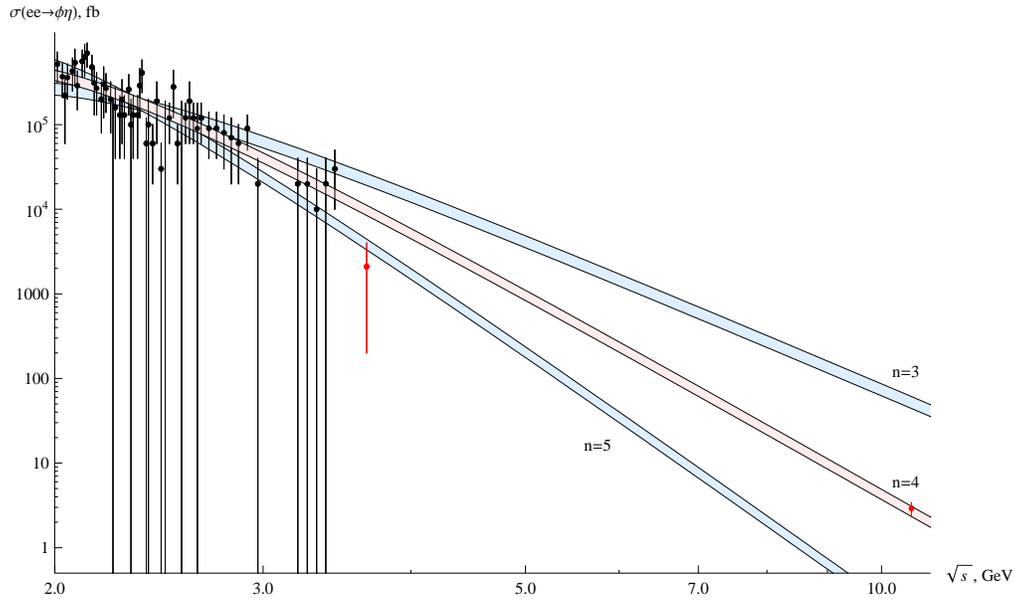}
\par\end{centering}
\caption{Different hypothesis on the energy dependence for the
$\sigma(e^{+}e^{-}\to\phi\eta)$. The constants $a_n$
are fixed from the experimental values of this cross section
in the low energy region ($2\,\mbox{GeV}\le\sqrt{s}\le3.5\,\mbox{GeV}$).\label{fig:PhiEta}}
\end{figure}

First we are going to consider the process $e^+e^- \to \phi \eta$. To fix the constants $a_{n}$ one can use  low energy data \cite{Aubert:2007ef}. When the constants $a_{n}$ for different hypothesis are fixed, one can use them to predict the cross section in the high energy region and then compare this prediction with available high energy data \cite{Aubert:2006xw} measured by BaBar collaboration
\begin{eqnarray*}
\sigma_\mathrm{BaBar}(e^+e^-\to\phi\eta) &=& 2.9\pm0.5\,\mathrm{fb}.
\end{eqnarray*}
Our results are shown in Fig. \ref{fig:PhiEta} and Table \ref{tab:params}. From these figure and table it can be clearly seen, that only $n=4$ describes satisfactory  low and high energy experimental BaBar data \cite{Aubert:2007ef, Aubert:2006xw}. This result is in agreement with the predictions of perturbative QCD. Form Fig. \ref{fig:PhiEta} one can see that low energy CLEO-c data \cite{Adam:2004pr} are in disagreement with hypothesis $n=3,4$. Only hypotheses $n=5$ does not contradict to the experimental results. However, if we assume that this hypotheses is correct we will faced with dramatic contradiction with the high energy BaBar data (see Fig. \ref{fig:PhiEta} and Table \ref{tab:params}). It should be also noted that in papers \cite{Lu:2007hr,Gerard:1997ym}  it was stated that the energy dependence $\sigma(e^+e^-\to\phi\eta)\sim 1/s^3$ describes experimental data more accurately. We believe  that the disagreement of this statement with our conclusion arises from the fact that the authors of papers \cite{Lu:2007hr,Gerard:1997ym} did not take into account low energy experimental result \cite{Aubert:2007ef}.

In view of this the question arises: is it possible to use asymptotic behavior of the cross section (\ref{asympt}) in the region $\sqrt s \in (2,3.5)$ GeV. First, one can estimate the cross uncertainty due to the power corrections as $\sim M^2/s \sim$. Even, for the heaviest meson $\phi$ and the smallest $\sqrt s$ from the region $\sqrt s \in (2,3.5)$ the error is $\sim 0.25 \%$. We can also determine the size of power corrections from the fitting data \cite{Aubert:2007ef} by the asymptotic form (\ref{asympt}) plus next-to-leading-order power correction. Our analysis shows that the uncertainty is not greater than $10\%\sim 20 \%$. We believe that these arguments confirm the applicability of the asymptotic expression for the cross section.

\begin{figure}
\begin{centering}
\includegraphics[scale=0.5]{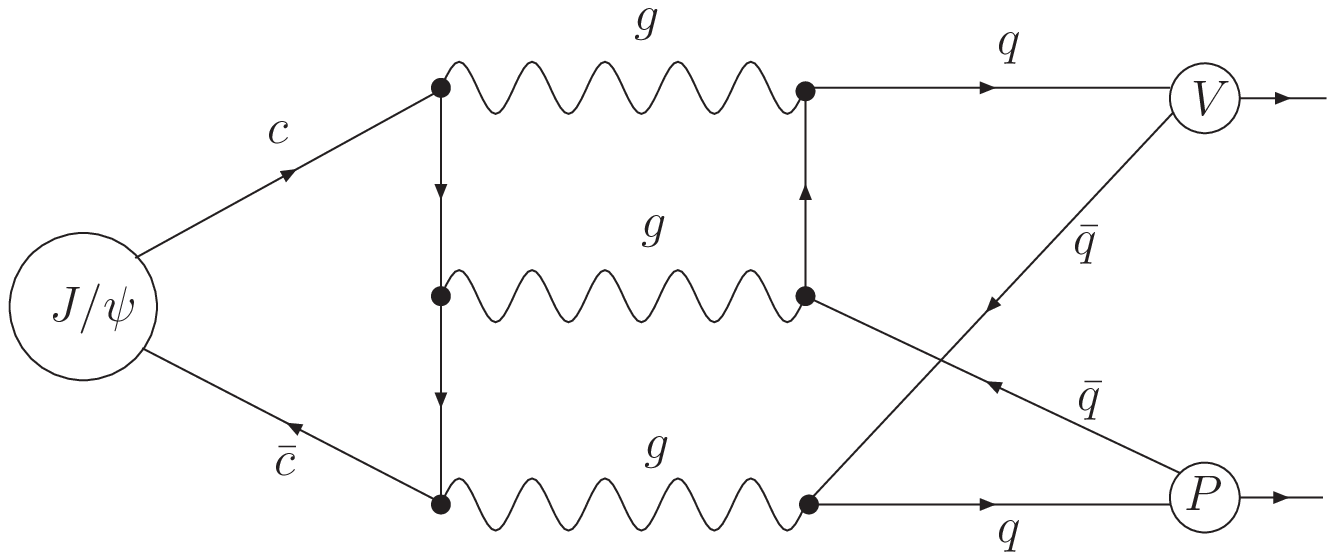}\includegraphics[scale=0.5]{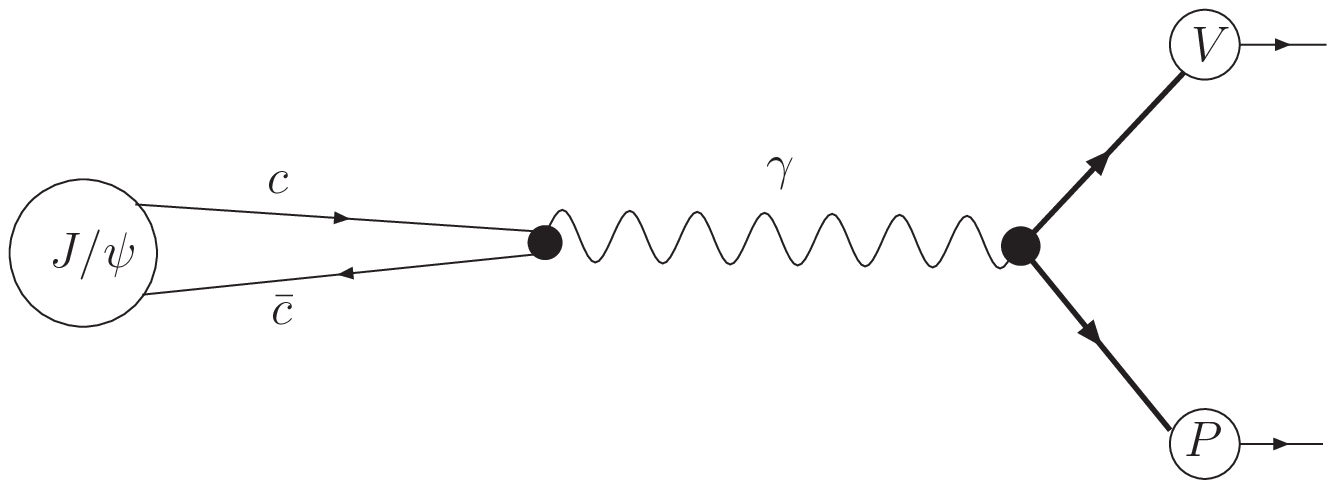}
\par\end{centering}
\caption{Typical diagrams for $J/\psi\to VP$ decay\label{diag:Psi}}
\end{figure}

\begin{figure}[p]
\begin{centering}
\includegraphics[scale=0.7]{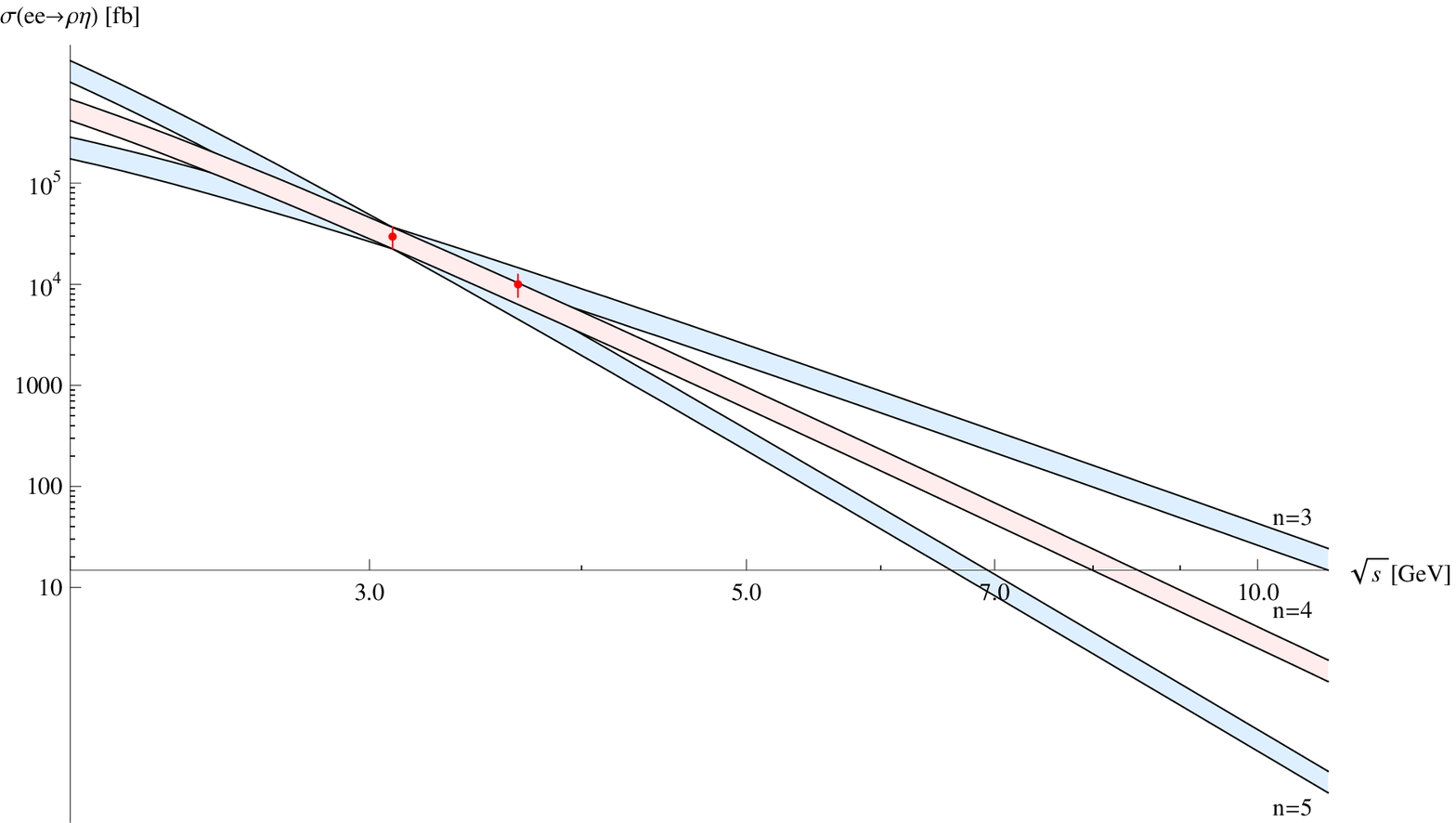}
\par\end{centering}
\caption{
Different hypothesis on the energy dependence for the
$\sigma(e^{+}e^{-}\to\rho\eta)$. The constants $a_n$
are fixed from the $J/\psi\to\rho\eta$ branching fraction (the
rightmost point represents the cross section at $\sqrt{s}=M_{J/\psi}$
calculated using relations (\ref{eq:SIGMA}), (\ref{eq:F}) and (\ref{eq:Br}).
) \label{fig:RhoEta}}
\end{figure}

\begin{figure}[p]
\begin{centering}
\includegraphics[scale=0.8]{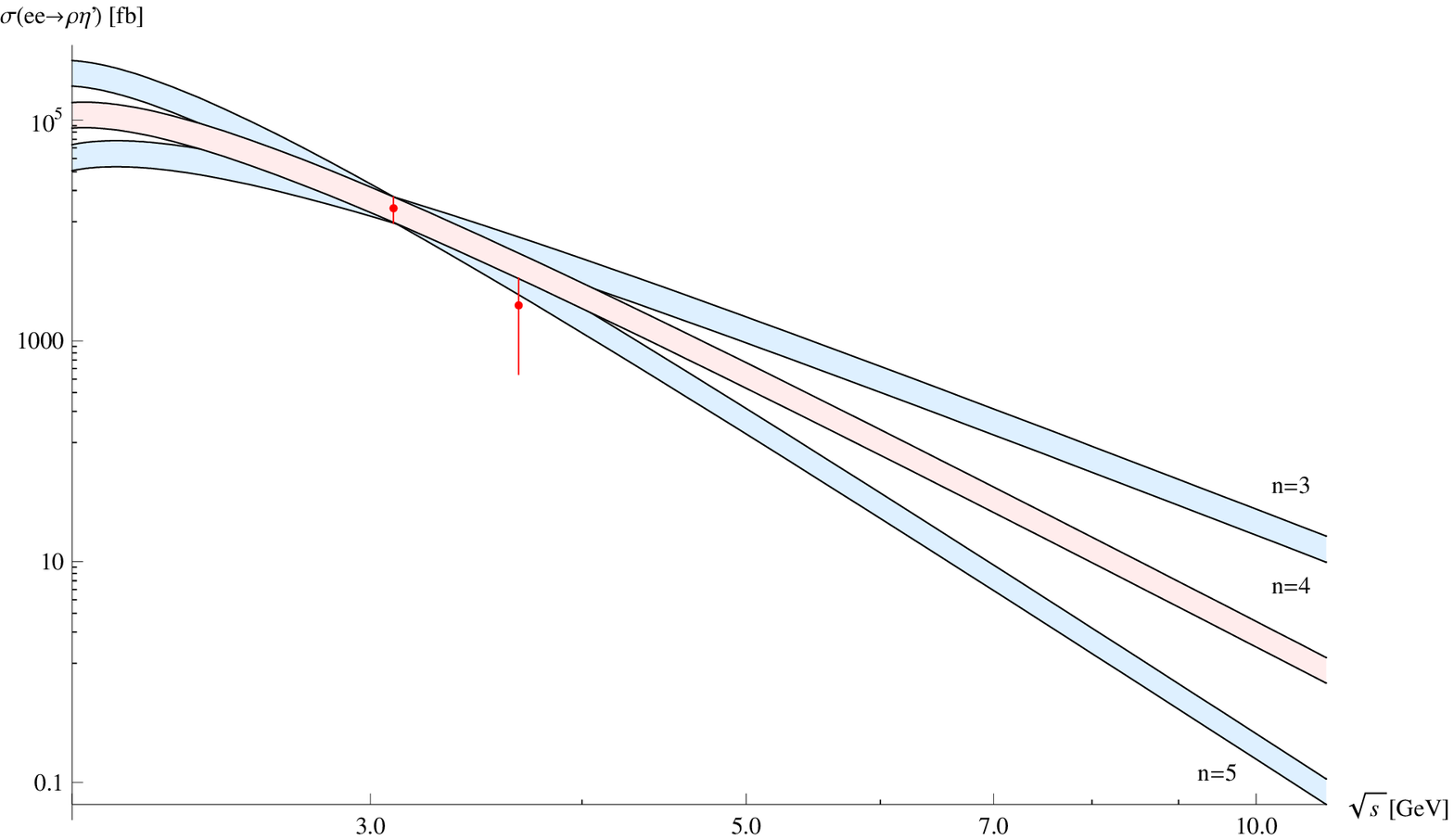}
\par\end{centering}
\caption{
Different hypothesis on the energy dependence for the
$\sigma(e^{+}e^{-}\to\rho\eta')$. The constants $a_n$
are fixed from the $J/\psi\to\rho\eta'$ branching fraction (the
rightmost point represents the cross section at $\sqrt{s}=M_{J/\psi}$
calculated using relations (\ref{eq:SIGMA}), (\ref{eq:F}) and (\ref{eq:Br}).
\label{fig:RhoEtaP}}
\end{figure}

Now let us consider the processes $e^{+}e^{-}\to\rho\eta$,  $e^{+}e^{-}\to\rho\eta'$. Unfortunately at the moment only the low energy CLEO-c data \cite{Adam:2004pr} are available for these processes. As it is seen from Tab. \ref{tab:params} the experimental error of these data are rather large. More precise experimental data can be obtained from  the decays $J/\psi\to\rho\eta, J/\psi\to\rho\eta'$. Corresponding branching fraction are equal to \cite{Yao:2006px}
\begin{eqnarray*}
\mbox{Br}\left(e^{+}e^{-}\to\rho\eta\right) & = & \left(1.93\pm0.23\right)\times10^{-4},\\
\mbox{Br}\left(e^{+}e^{-}\to\rho\eta'\right) & = & \left(1.05\pm0.18\right)\times10^{-4}.
\end{eqnarray*}
Generally speaking, these decays can proceed both via strong and electromagnetic interaction (see Fig. \ref{diag:Psi} for the typical diagrams). Because of the isospin violation, however, the gluon induced diagrams are strongly suppressed, and purely electromagnetic diagram gives the dominant contribution. The branching fraction of these decays are equal to
\begin{eqnarray}
\mbox{Br}(J/\psi\to VP) & = & \left(\frac{|p|}{M_{J/\psi}}\right)^{3}\left|\frac{a_{n}(M_{J/\psi}^{2})}{M_{J/\psi}^{n}}\right|^{2}M_{J/\psi}^{2}\mbox{Br}(J/\psi\to e^{+}e^{-}).\label{eq:Br}
\end{eqnarray}
From this relation we can determine the value of the function $a_{n}(M_{J/\psi}^{2}$) and, neglecting energy dependence in $a_{n}(s)$ (this assumptions leads to additional errors, which can be estimated as $\sim M^{2}/s\sim10\%$), predict the cross sections of the processes $e^{+}e^{-}\to\rho\eta$ and $e^{+}e^{-}\to\rho\eta'$ over the large energy region. The results are shown in Figs. \ref{fig:RhoEta}, \ref{fig:RhoEtaP} and Table \ref{tab:params}. From these results it is seen that the low energy data do not allow to understand what hypotheses gives the best agreement with the experiments. So, to determine the asymptotic behavior unambiguously the high energy data are needed. However, if we assume that the asymptotic behavior for the cross section of the process $e^+e^- \to \phi \eta$ is the same as that for the processes $e^{+}e^{-}\to\rho\eta$,  $e^{+}e^{-}\to\rho\eta'$, one can predict the cross sections of these processes at energies $\sqrt s =3.67, 10.6$ GeV
\begin{eqnarray}
\sigma_{\sqrt s=3.67~{GeV} } ( e^{+}e^{-}\to\rho\eta ) &=& 8 \pm 2~ \mbox{pb}, ~~~~~~~
\sigma_{\sqrt s=3.67~{GeV}  } ( e^{+}e^{-}\to\rho\eta' ) = 5 \pm 3~ \mbox{pb}, \nonumber \\
\sigma_{\sqrt s=10.6~{GeV}  } ( e^{+}e^{-}\to\rho\eta ) &=& 2.1 \pm 0.5~ \mbox{fb}, ~~~~~~
\sigma_{\sqrt s=10.6~{GeV}  } ( e^{+}e^{-}\to\rho\eta' ) = 1.4 \pm 0.4~ \mbox{fb}.
\end{eqnarray}
The cross sections at the energy $\sqrt s=3.67$ GeV are in agreement with the CLEO-c data.

\begin{figure}[p]
\begin{centering}
\includegraphics[scale=0.7]{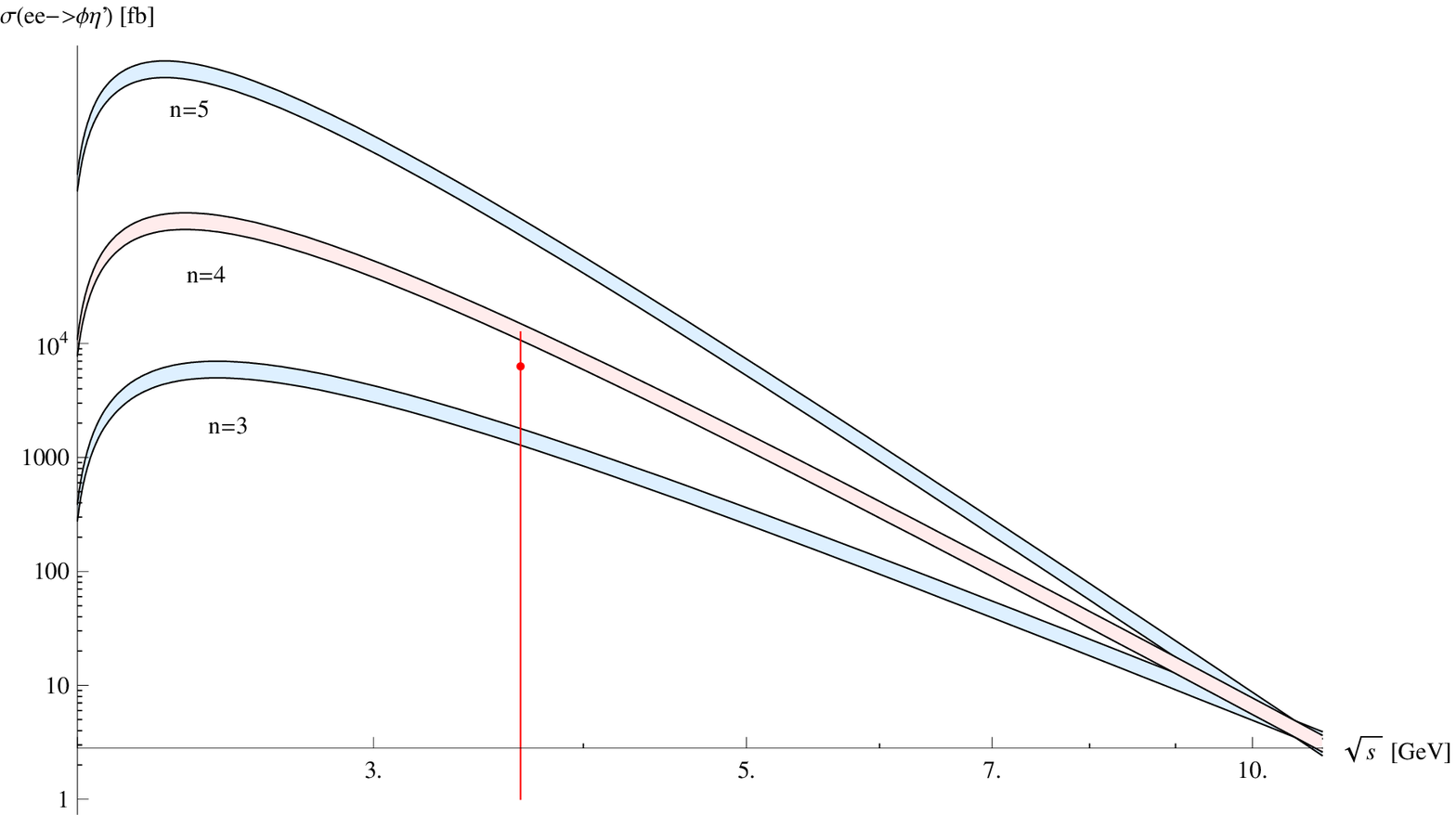}
\par\end{centering}
\caption{
Different hypothesis on the energy dependence for the
$\sigma(e^{+}e^{-}\to\phi\eta')$. The constants $a_n$
are fixed from the experimental values of this cross section
at $\sqrt{s}=10.6$ GeV.
\label{fig:PhiEtaP}}
\end{figure}

Let us now proceed with the reaction $e^{+}e^{-}\to\phi\eta'$. In this case experimental information is even more poor. In the case one cannot use the branching fraction of the decay $\mbox{Br} (J/\Psi \to \phi \eta')$, since in this case  isospin is conserved and the diagram shown in Fig. \ref{diag:Psi}a gives the main contribution to $J/\psi\to\phi\eta'$ decay. In addition, CLEO-c gives us only upper bound on the cross section $\sigma(e^{+}e^{-}\to\phi\eta')$, and no experimental information in high energy region is available. Using $\eta-\eta'$ mixing it is possible, however, to estimate the cross section $\sigma(e^+e^-\to\phi\eta')$ from the value of the $e^+e^-\to\phi\eta$ cross section. The mixing of pseudoscalar mesons can be described by different parameterizations \cite{Feldmann:1999uf}. In our paper we will use the parametrization of $\eta-\eta'$ mixing in quark flavor basis with one mixing angle \cite{Feldmann:1998vh}:
\begin{eqnarray}
\left(\begin{array}{c}
\eta\\
\eta'\end{array}\right) & = & \left(\begin{array}{cc}
\cos\Phi & -\sin\Phi\\
\sin\Phi & \cos\Phi\end{array}\right)\left(\begin{array}{c}
\eta_{n}\\
\eta_{s}\end{array}\right),
\label{eq:mixing}
\end{eqnarray}
where $\eta_{n}=(u\bar{u}+d\bar{d})/\sqrt{2}$ and $\eta_{s}=s\bar{s}$ represents the basis of the quark mixing scheme. Using this mixing scheme it is easy to obtain the following relation between $e^+e^-\to\phi\eta$ and $e^+e^-\to\phi\eta'$ cross sections:
\begin{eqnarray*}
\frac{\sigma(e^+e^-\to\phi\eta')}{\sigma(e^+e^-\to\phi\eta)} &=& \cot^2\Phi.
\end{eqnarray*}

There are plenty of theoretical and experimental works dedicated to the determination of the mixing angle. The well known estimation, based on Gell-Mann-Okubo mass formulas give the value of $\Phi$ about $32^o$ for linear mass formula and about $45^o$ for a quadratic case. The analysis of the axial anomaly generated decays $\eta, \eta'\to\gamma\gamma$ was performed in papers \cite{Donoghue:1986wv,Gilman:1987ax} and the estimate $\Phi= 30^o\div 35^o$ was obtained. In papers \cite{Akhoury:1987ed,Ball:1995zv} another anomaly based investigation of a large set of decay processes was performed and the value $\Phi=38^o\pm2^0$ was presented. The authors of the recent work \cite{Klopot:2008ec} used dispersive approach to $\eta-\eta'$ mixing and obtained the value
\begin{eqnarray*}
\Phi &=& 39.4^o\pm 1^o.
\end{eqnarray*}
It is interesting to note that this value is close to phenomenological value $\Phi=39.3^o\pm1^o$, presented in pioneering work \cite{Feldmann:1998vh}. We will use this value of the mixing angle in our article. The error of this angle is small in comparison with the error in $e^+e^-\to\phi\eta$ cross section.

Knowing the value of the cross section $\sigma(e^+e^-\to\phi\eta)$ one can determine the
value of the cross section $\sigma(e^+e^-\to\phi\eta')$ at the energy $\sqrt s=10.6$ GeV.
\begin{eqnarray*}
\sigma_{\sqrt s=10.6~{GeV} }(e^+e^-\to\phi\eta') = 4.2 \pm 0.7~ \mbox{fb} .
\end{eqnarray*}
From this we can calculate the value of the constants $a_n$ for different hypothesis and predict the cross section of the process $e^+e^-\to\phi\eta'$ in the low energy region. We show the energy dependence of this cross section in Fig. \ref{fig:PhiEtaP}. From this figure one sees that the value $n=5$ contradicts experimental results at $\sqrt{s}=3.67$ GeV, while $n=3$ and $n=4$ do not. From theoretical arguments and presented above analysis of $e^+e^-\to\phi\eta$ reaction we think, that the value $n=4$ is more preferable. For this hypothesis we can predict the value of the cross section at energy $\sqrt s=3.67$ GeV
\begin{eqnarray*}
\sigma_{\sqrt s=3.67~{GeV} }(e^+e^-\to\phi\eta') = 12.8 \pm 2.1~ \mbox{pb},
\end{eqnarray*}
which does not contradict to the CLEO-c data.

In table \ref{tab:params} we present the results of this section. Second column contains constants $a_{3,4,5}$ for different final states, obtained from low energy fits. In the third and fifth columns experimental results for the cross sections $\sigma(e^{+}e^{-}\to VP)$ at the center mass energies $\sqrt{s}=3.67$ GeV and 10.6 GeV are presented. The results of the calculation  are shown in the forth and sixth columns. From this table it is clear, that only relation $\sigma\sim1/s^{4}$ agrees with experiment.

\begin{table*}
\begin{centering}
\begin{tabular}{|c|c||c|c||c|c|}
\hline
 &  & \multicolumn{2}{c||}{$\sigma(\sqrt{s}=3.67\,\mbox{GeV})$ pb} &
 \multicolumn{2}{c|}{$\sigma(\sqrt{s}=10.6\,\mbox{GeV})$ fb}\tabularnewline
\hline
$VP$  & parrams  & exp  & fit results  & exp & fit results\tabularnewline
\hline
\hline
 & $a_{3}=1.8\pm0.2\,\mbox{GeV}^{2}$  &  & $12\pm3$  & & $24\pm6$ \tabularnewline
$\rho\eta$  & $a_{4}=5.6\pm0.7\,\mbox{GeV}^{3}$  & $10\pm2.5$  & $8\pm2$ & -- & $2.1\pm0.5$\tabularnewline
 & $a_{5}=17.3\pm2.1\,\mbox{GeV}^{4}$  &  & $6\pm1$ &   & $0.18\pm0.04$ \tabularnewline
\hline
 & $a_{3}=1.5\pm0.2\,\mbox{GeV}^{2}$  &  & $7\pm2$ & & $17\pm4$ \tabularnewline
$\rho\eta^{'}$  & $a_{4}=4.7\pm0.6\,\mbox{GeV}^{3}$  & $2.1\pm1.6$  & $5\pm3$ & -- & $1.4\pm0.4$\tabularnewline
 & $a_{5}=14.5\pm1.9\,\mbox{GeV}^{4}$  &  & $3.5\pm0.9$ & & $0.12\pm0.03$\tabularnewline
\hline
 & $a_{3}=2.7\pm0.2\,\mbox{GeV}^{2}$  &  & $23\pm4$ & & $52\pm8$\tabularnewline
$\phi\eta$  & $a_{4}=6.4\pm0.4\,\mbox{GeV}^{3}$  & $2.1\pm1.9$  & $9.8\pm1.3$ & $2.9\pm0.5$ & $2.7\pm0.4$\tabularnewline
 & $a_{5}=14.8\pm1.1\,\mbox{GeV}^{4}$  &  & $3.8\pm0.5$ & & $0.13\pm0.02$\tabularnewline
\hline
 & $a_{3}=0.76\pm0.06\,\mbox{GeV}^{2}$  &  & $1.5\pm0.2$ & & $4.2\pm0.7$\tabularnewline
$\phi\eta'$  & $a_{4}=8.1\pm0.7\,\mbox{GeV}^{3}$  & $<12.6$  & $12.8\pm2.1$ & -- & $4.2\pm0.7$\tabularnewline
 & $a_{5}=85.\pm7.\,\mbox{GeV}^{4}$  &  & $110\pm18$  & & $4.2\pm0.7$\tabularnewline
\hline
\end{tabular}
\par\end{centering}

\caption{The constants $a_n$ and the cross sections at $\sqrt{s}=3.67$ and 10.6 GeV in comparison with the experimental data\label{tab:params}. The second column contains the constants $a_{3,4,5}$ for the different final states, obtained from the low energy data. In the third and fifth  columns the experimental results for the cross sections  $\sigma(e^{+}e^{-}\to VP)$ at the center mass energy $\sqrt{s}=3.67$ GeV and $\sqrt{s}=10.6$ GeV  are presented. The results of the calculation  are shown in the forth and sixth columns.
}
\end{table*}

\section{Calculation of the cross sections.\label{sec:Light-Cone}}

\subsection{Numerical parameters and distribution amplitudes.}

Now let us try to estimate the cross sections of the processes studied in the last section theoretically. To calculate the cross sections of the processes $e^{+}e^{-}\to VP$ for $V=\phi,\rho,P=\eta,\eta'$ one needs to know distribution amplitudes (DA) of final mesons. For the vector mesons the DAs needed in the calculation can be written as \cite{Chernyak:1983ej,Bondar:2004sv}
\begin{eqnarray}
{\langle V_{\lambda}(p)|{\bar{q}}_{\beta}(z)\, q_{\alpha}(-z)|0\rangle}_{\mu}=\frac{f_{V}M_{V}}{4}\int_{o}^{1}dx\, e^{i(pz)(2x-1)}\biggl\{{\widehat{p}}\,\frac{(e_{\lambda}z)}{(pz)}\, V_{L}(x)+\biggl({\widehat{e}}_{\lambda}-{\widehat{p}}\,\frac{(e_{\lambda}z)}{(pz)}\biggr)\, V_{\perp}(x)+\nonumber \\
\frac{f^{T}(\mu)}{f_{V}M_{V}}(\sigma_{\mu\nu}e_{\lambda}^{\mu}\, p^{\nu})\, V_{T}(x)+\frac{1}{2}(\epsilon_{\mu\nu\alpha\beta}\gamma_{\mu}\gamma_{5}\, e_{\lambda}^{\nu}\, p^{\alpha}z^{\beta})\, V_{A}(x)\biggl\}_{\alpha\beta},\label{vector}
\end{eqnarray}
where $x$ is the fraction of momentum carried by quark, $f_{V},f_{T}(\mu),M_{V}$ are the leptonic, tensor constants and the mass of vector meson. In the calculation $\phi$ meson is assumed to be composed of $s$ quarks, so in (\ref{vector}) $q=s$. For $\rho^{0}$ meson isospin 1 combination $\bar{q}q=(\bar{u}u-\bar{d}d)/\sqrt{2}$ is assumed. The models for DAs and the decay constants $f_{V},f_{T}(\mu)$ for $\phi$ and $\rho$ mesons will be taken from paper \cite{Ball:1998sk}.

In the framework of the  quark mixing scheme (\ref{eq:mixing}) the decay constants
\begin{eqnarray*}
\langle P(p)|\bar{n}\gamma^{\mu}\gamma_{5}n|0\rangle & = & if_{P}^{n}p^{\mu},\\
\langle P(p)|\bar{s}\gamma^{\mu}\gamma_{5}s|0\rangle & = & if_{P}^{s}p^{\mu},\end{eqnarray*}
 needed in the calculation can be expressed through the constants
\begin{eqnarray*}
\langle\eta_{n}(p)|\bar{n}\gamma^{\mu}\gamma_{5}n|0\rangle & = & if^{n}p^{\mu},\\
\langle\eta_{s}(p)|\bar{s}\gamma^{\mu}\gamma_{5}s|0\rangle & = & if^{s}p^{\mu},\end{eqnarray*}
 as follows \begin{eqnarray*}
\left(\begin{array}{cc}
f_{\eta}^{n} & f_{\eta}^{s}\\
f_{\eta'}^{n} & f_{\eta'}^{s}\end{array}\right) & = & \left(\begin{array}{cc}
\cos\Phi & -\sin\Phi\\
\sin\Phi & \cos\Phi\end{array}\right)\left(\begin{array}{cc}
f^{n} & 0\\
0 & f^{s}\end{array}\right).\end{eqnarray*}
 In turn, the constants $f^{n},f^{s}$ and the mixing angle $\Phi$
can be determined from experiment \cite{Feldmann:1998vh} \begin{eqnarray*}
f^{n} & = & (1.07\pm0.02)f_{\pi},\qquad f^{s}=(1.34\pm0.06)f_{\pi}.
\end{eqnarray*}
Within this mixing pattern the DAs needed in the calculation can be written
in the following form \cite{Chernyak:1983ej,Bondar:2004sv} \begin{eqnarray*}
{\langle P(p)|{\bar{n}}_{\beta}(z)\, n_{\alpha}(-z)|0\rangle}_{\mu} & = & i\frac{f_{P}^{n}{M_{P}}}{4}\int_{0}^{1}dye^{i(pz)(2y-1)}\biggl\{\frac{\hat{p}\,\gamma_{5}}{M_{P}}\, P_{A}^{n}(y)-f_{p}^{n}(\mu)\,\gamma_{5}\, P_{P}^{n}(y)\biggr\},\\
{\langle P(p)|{\bar{s}}_{\beta}(z)\, s_{\alpha}(-z)|0\rangle}_{\mu} & = & i\frac{f_{P}^{s}{M_{P}}}{4}\int_{0}^{1}dye^{i(pz)(2y-1)}\biggl\{\frac{\hat{p}\,\gamma_{5}}{M_{P}}\, P_{A}^{s}(y)-f_{p}^{s}(\mu)\,\gamma_{5}\, P_{P}^{s}(y)\biggr\},\end{eqnarray*}
 where \begin{eqnarray*}
f_{p}^{n}(\mu) & = & \frac{1}{2m_{n}(\mu)}\biggl[m_{\eta}^{2}\cos^{2}\Phi+m_{\eta'}^{2}\sin^{2}\Phi-\frac{\sqrt{2}f^{s}}{f^{n}}(m_{\eta'}^{2}-m_{\eta}^{2})\cos\Phi\sin\Phi\biggr],\\
f_{p}^{s}(\mu) & = & \frac{1}{2m_{s}(\mu)}\biggl[m_{\eta'}^{2}\cos^{2}\Phi+m_{\eta}^{2}\sin^{2}\Phi-\frac{f^{n}}{\sqrt{2}f^{s}}(m_{\eta'}^{2}-m_{\eta}^{2})\cos\Phi\sin\Phi\biggr].\end{eqnarray*}
The calculation will be done with the masses $m_{s}(1~\mbox{GeV})=150$ MeV, $m_{n}(1\mbox{GeV})=(m_{u}(1~\mbox{GeV})+m_{d}(1~\mbox{GeV}))/2=5$ MeV and it will be assumed that $P_{A}^{s}(y)=P_{A}^{n}(y)=P_{A}(y)$ and $P_{P}^{s}(y)=P_{P}^{n}(y)=P_{P}(y)$. The models of the leading twist DAs $P_A$ for the $\eta$ and $\eta'$ mesons will be taken from paper \cite{Kroll:2002nt}. For the function $P_{P}(y)$ the asymptotic form will be used.

It should be noted that the wave functions and the constants introduced above depend on renormalization scale $\mu$. Our calculation shows that the scale dependence of the DAs is not very important and below it will be ignored. At the same time the scale dependence of the constants is important and it will be taken into account. The calculation will be done at the scale $\mu=\sqrt{s}/2$.

\subsection{Numerical results and discussion.}

Having introduced the designations of the DAs one can proceed with the calculation of the cross sections. First we are going to consider the diagrams similar to that shown in Fig.\ref{fig:diags}a. It is not difficult to calculate the contribution of these diagrams to the formfactor $F(s)$
\begin{eqnarray*}
|F(s)| & = & \frac{4\pi\alpha}{s}\frac{f_{V}}{M_{V}}\biggl(\frac{e_{u}^{2}+e_{d}^{2}}{\sqrt{2}}f_{P}^{n}+e_{s}^{2}f_{P}^{s}\biggr)\int_{0}^{1}\frac{dy}{y_{1}y_{2}}P_{A}(y),
\end{eqnarray*}
where $e_{u},e_{d},e_{s}$ are the charges of $u,d,s$ quarks correspondingly. For the process $e^{+}e^{-}\to\phi\eta$ at the energy $\sqrt{s}=10.6$ GeV we have $\sigma=0.01$ fb, what is by two orders of magnitude less than the experimental result. Form this it is clear that the contribution of the diagrams shown in Fig.\ref{fig:diags}a is negligible and below it will be ignored. It should be noted that in the limit $s\to\infty$ the contribution of Fig.\ref{fig:diags}a diagrams to the cross section has the following behavior $\sim1/s^{2}$. So, at extremely large energy this diagrams give the dominant contribution. Our calculation shows that this happens at energies $\sqrt{s}\geq35$ GeV.

Now we are going to consider the other diagrams shown in Fig. \ref{fig:diags}. The leading asymptotic behavior of these diagrams is $\sigma\sim1/s^{4}$. First, it should be noted that the contribution of the diagrams shown in Fig. \ref{fig:diags}b is very small \cite{Lu:2007hr} and it will be ignore below. This fact results from rather small admixture of $|GG\rangle$ fock state in the pseudoscalar mesons $\eta,\eta'$. Now we proceed to the calculation of the diagrams shown in Fig. \ref{fig:diags}c. The contribution of these diagrams to the  $F(s)$ for the processes under study can be written as follows \cite{Bondar:2004sv}
\begin{eqnarray}
|F_{\phi\eta}(s)| & = & \frac{32\pi}{9}\frac{f_{V}f^{s}M_{V}\sin\Phi}{s^{2}}~e_{s}I_{0},\nonumber \\
|F_{\phi\eta'}(s)| & = & \frac{32\pi}{9}\frac{f_{V}f^{s}M_{V}\cos\Phi}{s^{2}}~e_{s}I_{0},\nonumber \\
|F_{\rho\eta}(s)| & = & \frac{32\pi}{9}\frac{f_{V}f^{n}M_{V}\cos\Phi}{2s^{2}}~(e_{u}-e_{d})I_{0},\nonumber \\
|F_{\rho\eta'}(s)| & = & \frac{32\pi}{9}\frac{f_{V}f^{n}M_{V}\sin\Phi}{2s^{2}}~(e_{u}-e_{d})I_{0}.\label{res}
\end{eqnarray}
In the above expressions
\begin{eqnarray}
I_{0} & = & \int_{0}^{1}dx\int_{0}^{1}dy~\alpha_{s}(\mu)\left\{ \frac{f_{t}(\mu)}{M_{V}}f_{p}(\mu)\frac{V_{T}(x)P_{P}(y)}{x^{2}\, y}+\right.\label{I0res}\\
 & + & \frac{1}{2}\frac{V_{L}(x)\, P_{A}(y)}{x\, y}+(1-2y)\frac{V_{\perp}(x)\, P_{A}(y)}{x\, y^{2}}+\left.\frac{1}{8}\,\frac{(1+y)V_{A}(x)P_{A}(y)}{x^{2}\, y^{2}}\right\} ,\nonumber
\end{eqnarray}
 where $f_{t}(\mu)=f_{T}(\mu)/f_{V}$, $f_{p}(\mu)=f_{p}^{s}(\mu)$ for the processes with $\phi$ meson in the final state and $f_{p}(\mu)=f_{p}^{n}(\mu)$ for the processes with $\rho$ meson in the final state.

\begin{table}
\begin{centering}
\begin{tabular}{|c|c|c|c||c|c|c|}
\hline
 & \multicolumn{3}{c||}{$\sigma(\sqrt{s}=3.67\mbox{ GeV})$ pb} & \multicolumn{3}{c|}{$\sigma(\sqrt{s}=10.6\mbox{ GeV})$ fb}\tabularnewline
\hline
$VP$  & exp \cite{Adam:2004pr}  & \cite{Lu:2007hr}  & this work & exp \cite{Aubert:2006xw}  & \cite{Lu:2007hr}  & this wok\tabularnewline
\hline
\hline
$\rho\eta$  & $10\pm2.5$  & $8.1\div16.6$  & $3.7\div6.3$ & -- & $2.4\div3.1$  & $2.4 \div 3.5$ \tabularnewline
\hline
$\rho\eta'$  & $2.1\pm1.6$  & $4.3\div8.6$  & $2.1\div3.6$ & --  & $1.5\div2.1$  & $1.6 \div 2.3$ \tabularnewline
\hline
$\phi\eta$  & $2.1\pm1.9$  & $9.6\div19.1$  & $ 3.7\div6.1$ & $2.9\pm0.5$ & $3.3\div4.3$  & $2.4 \div 3.4$ \tabularnewline
\hline
$\phi\eta'$  & $<12.6$  & $11.5\div22.6$  & $4.6 \div7.6$ & -- & $4.4\div5.8$  & $3.5 \div 5.0 $ \tabularnewline
\hline
\end{tabular}
\par\end{centering}
\caption{The results of the calculation. The second and fifth columns contain experimental results for the cross sections at energies $\sqrt{s}=3.67$ GeV and $\sqrt{s}=10.6$ GeV correspondingly. The result obtained in paper \cite{Lu:2007hr} are shown in the third and sixth columns. The results obtained in this paper are shown in the fourth and seventh columns.\label{res1}
}
\end{table}

The following point deserves consideration. The models for the DAs that we use in our calculation are truncated series in Gegebauer polynomials. This means that the end point behavior ($x\to0,1$) of these DAs coincides with the end point behavior of the corresponding asymptotic DAs. From this it is not difficult to see that the integral $I_{0}$ is logarithmically divergent. One way to regularize this divergence is to introduce cut off parameter $x_{0}$. In our calculation we take the following value of cut off parameter $x_{0}=\Lambda/\sqrt{s}$. $\Lambda$ is of order of typical hadronic scale, which is of order of several hundreds MeV. Physical meaning of this cut off can be understood as follows: if $x\sim x_{0}$ quark momentum is of order of $\sim\Lambda$ and in this region one must take into account transverse motion in hadron what regularizes the whole integral $I_{0}$. The parameter $\Lambda$ can be determined from  available experimental results. We determined this parameter from the cross section of the process $e^+ e^- \to \phi \eta$ at $\sqrt s = 10.6$ GeV. Thus we get $\Lambda = 130^{+25}_{-17}$ MeV. The variation of parameter $\Lambda$ corresponds $1 \sigma$ deviation from the central value measured at Belle collaboration.

The results of the calculation are shown in Table {\ref{res1}}. The second and fifth columns contain the experimental results for the cross sections at the energies $\sqrt{s}=3.67$ GeV and $\sqrt{s}=10.6$ GeV correspondingly. The result obtained in paper \cite{Lu:2007hr} are shown in the third and sixth columns. The results obtained in this paper are shown in the fourth and seventh columns. The variation of the results are due to the variation in parameter $\Lambda$. It is seen from this table that  the results of the calculation are in satisfactory agreement with the experiment.

It should be noted here that in addition to the diagrams shown in Fig. \ref{fig:diags}, there is additional contribution to the formfactor $F(s)$ which was not considered in this paper. This contribution appears if one takes into account higher fock state of the vector and pseudoscalar mesons $|q\bar{q}G\rangle$ and it's asymptotic behavior is also $\sigma\sim1/s^{4}$. This contribution was considered in papers \cite{Chernyak:1983ej,Gorsky:1985aw}. In paper \cite{Chernyak:1983ej} the authors asserted that the fock state $|q\bar{q}G\rangle$ gives very important contribution to the cross section and must be taken into account. Contrary to this conclusion, the author of paper \cite{Gorsky:1985aw} asserted that the main contribution arises from the diagrams shown in Fig. \ref{fig:diags}c. So, the question about the role of higher fock state $|q\bar{q}G\rangle$ in the total cross section deserves separate consideration and it will not be considered here.

\section{Discussion and conclusion.}

In this work we have studied the production of light mesons $\rho\eta$, $\rho\eta'$, $\phi\eta$ and $\phi\eta'$ in the high energy electron-positron annihilation.

The first question studied in this paper is the asymptotic behavior of the cross sections of the processes under consideration. Perturbative QCD predicts, that the cross section of the reaction $e^{+}e^{-}\to VP$ has the following asymptotic behavior $\sigma\sim1/s^{4}$ in the limit $s \to \infty$. Experimental measurement of the cross section of the process $e^+ e^- \to \phi\eta$  at the large center mass energy $\sqrt s=10.6$ GeV \cite{Aubert:2006xw} and the low energy experimental data $\sqrt s \sim 2-4$ GeV  \cite{Aubert:2007ef}  give us the possibility to study the cross section in the broad energy region. As the result, we have determined the asymptotic behavior of the cross section of $e^+ e^- \to \phi\eta$ in the limit $s \to \infty$, which is in agreement with perturbative QCD prediction. As to the other processes under study, there are no high energy experimental data, which allow us to confirm perturbative QCD prediction for these processes. We would like to stress here that the high energy experimental data turned out to be crucial in the determination of the asymptotic behaviour of the cross sections. Assuming that the asymptotic behavior predicted by perturbative QCD is valid for the other processes under consideration, we have calculated the cross sections of the processes $e^+ e^- \to \rho\eta, \rho\eta', \phi\eta, \phi\eta'$ at the energies $\sqrt s= 3.67, 10.6$ GeV.

In addition, we have applied perturbative QCD approach to calculate the cross sections of the processes under study at the energies $\sqrt s = 3.67$ GeV and $\sqrt s = 10.6$ GeV. The results of this calculation are in satisfactory agreement with available experimental data.

The authors would like to thank A.A. Sokolov and M.M. Shapkin for useful and stimulating discussions. This work was partially supported by Russian Foundation of Basic Research under grant 07-02-00417. The work of V. Braguta was partially supported by  CRDF grant Y3-P-11-05 and president grant MK-2996.2007.2. The work of A. Luchinsky was partially supported by  president grant MK-110.2008.2 and Russian Science Support Foundation.


\begin{thebibliography}{10}
\bibitem{Lepage:1980fj}G.~P.~Lepage and S.~J.~Brodsky, Phys.\ Rev.\ D
\textbf{22}, 2157 (1980).

\bibitem{Chernyak:1983ej} V.~L.~Chernyak and A.~R.~Zhitnitsky,
Phys.\ Rept.\ \textbf{112}, 173 (1984).

\bibitem{Chernyak:1977fk} V.~L.~Chernyak, A.~R.~Zhitnitsky and
V.~G.~Serbo, %``Asymptotic hadronic form-factors in quantum chromodynamics,''
 JETP Lett.\ \textbf{26}, 594 (1977) {[}Pisma Zh.\ Eksp.\ Teor.\ Fiz.\ \textbf{26},
760 (1977)]. %%CITATION = ZFPRA,26,760;%%


\bibitem{Chernyak2} V.~L.~Chernyak and A.~R.~Zhitnitsky, %``Asymptotics Of Hadronic Form-Factors In The Quantum Chromodynamics. (In
 %Russian),''
 Sov.\ J.\ Nucl.\ Phys.\ \textbf{31}, 544 (1980) {[}Yad.\ Fiz.\ \textbf{31},
1053 (1980)]. %%CITATION = YAFIA,31,1053;%%


\bibitem{Chernyak3} V.~L.~Chernyak and A.~R.~Zhitnitsky, %``Asymptotic Behavior Of Hadron Form-Factors In Quark Model. (In Russian),''
 JETP Lett.\ \textbf{25}, 510 (1977) {[}Pisma Zh.\ Eksp.\ Teor.\ Fiz.\ \textbf{25},
544 (1977)]. %%CITATION = ZFPRA,25,544;%%


\bibitem{Abe:2002rb} K.~Abe \textit{et al.} {[}Belle Collaboration],
Phys.\ Rev.\ Lett.\ \textbf{89}, 142001 (2002) {[}arXiv:hep-ex/0205104].

\bibitem{Aubert:2005tj} B.~Aubert \textit{et al.} {[}BABAR Collaboration],Phys.\ Rev.\ D
\textbf{72}, 031101 (2005) {[}arXiv:hep-ex/0506062].

\bibitem{Chao1} K.~Y.~Liu, Z.~G.~He and K.~T.~Chao, %``Problems of double charm production in e+ e- annihilation at s**(1/2) =
 %10.6-GeV. ((V)),''
 Phys.\ Lett.\ B \textbf{557}, 45 (2003) {[}arXiv:hep-ph/0211181].

\bibitem{Chao2} K.~Y.~Liu, Z.~G.~He and K.~T.~Chao, %``Search for excited charmonium states in e+ e- annihilation at s**(1/2)  =
 %10.6-GeV,''
 {[}arXiv:hep-ph/0408141].

\bibitem{Chao3} Y.~J.~Zhang, Y.~j.~Gao and K.~T.~Chao, %``Next-to-leading order QCD correction to e+ e- --> J/psi + eta/c at
 %s**(1/2) = 10.6-GeV,''
 Phys.\ Rev.\ Lett.\ \textbf{96}, 092001 (2006) {[}arXiv:hep-ph/0506076];
%%CITATION = HEP-PH 0506076;%%


\bibitem{Chao4} Z.~G.~He, Y.~Fan and K.~T.~Chao, %``Relativistic corrections to $J/\psi$ exclusive and inclusive double   charm
 %production at B factories,''
 Phys.\ Rev.\ D \textbf{75}, 074011 (2007) {[}arXiv:hep-ph/0702239].

\bibitem{Choi:2007ze} H.~M.~Choi and C.~R.~Ji, %``Perturbative QCD analysis of exclusive $J/\psi+\eta_c$ production in
 %$e^+e^-$ annihilation,''
 Phys.\ Rev.\ D \textbf{76}, 094010 (2007) {[}arXiv:0707.1173 {[}hep-ph]].
%%CITATION = PHRVA,D76,094010;%%


\bibitem{Bondar:2004sv} A.~E.~Bondar and V.~L.~Chernyak, % ``Is the BELLE result for the cross section sigma(e+ e- --> J/psi +  eta/c) a
 %real difficulty for QCD?,''
 Phys.\ Lett.\ B \textbf{612}, 215 (2005) {[}arXiv:hep-ph/0412335].
%%CITATION = HEP-PH 0412335;%%



%\cite{Braguta:2005kr}
\bibitem{Braguta:2005kr} V.~V.~Braguta, A.~K.~Likhoded and A.~V.~Luchinsky,
%``Excited charmonium mesons production in e+ e- annihilation at s**(1/2) =
 %10.6-GeV,''
 Phys.\ Rev.\ D \textbf{72}, 074019 (2005) {[}arXiv:hep-ph/0507275].
%%CITATION = HEP-PH 0507275;%%


\bibitem{Braaten:2002fi} E.~Braaten and J.~Lee, %``Exclusive double-charmonium production in e+ e- annihilation,''
 Phys.\ Rev.\ D \textbf{67}, 054007 (2003) {[}arXiv:hep-ph/0211085];


%\cite{Bodwin:2007ga}
\bibitem{Bodwin:2007ga} G.~T.~Bodwin, J.~Lee and C.~Yu, %``Resummation of Relativistic Corrections to e+ e- -> J/psi+eta_c,''
 arXiv:0710.0995 {[}hep-ph]. %%CITATION = ARXIV:0710.0995;%%



%\cite{Ma:2004qf}
\bibitem{Ma:2004qf} J.~P. Ma and Z.~G.Si, Phys.\ Rev.\ D \textbf{70},
074007 (2004), {[}arXiv:hep-ph/0405111].



%\cite{Lu:2007hr}
\bibitem{Lu:2007hr} C.~D.~Lu, W.~Wang and Y.~M.~Wang, %``Exclusive processes $e^+e^-\to VP$ in $k_T$ factorization,''
 Phys.\ Rev.\ D \textbf{75}, 094020 (2007) {[}arXiv:hep-ph/0702085].
%%CITATION = PHRVA,D75,094020;%%





%\cite{Gerard:1997ym}
\bibitem{Gerard:1997ym} J.~M.~Gerard and G.~Lopez Castro, %``Tests for the asymptotic behaviour of the gamma* --> gamma pi0 form
 %factor,''
 Phys.\ Lett.\ B \textbf{425}, 365 (1998) {[}arXiv:hep-ph/9709404].
%%CITATION = PHLTA,B425,365;%%

%\cite{Aubert:2007ef}
\bibitem{Aubert:2007ef}
  B.~Aubert {\it et al.}  [BABAR Collaboration],
  %``The $e^+ e^-\to 2(\pi^+\pi^-)\pi^0$, 2(\pi^+\pi^-)\eta$, $K^+
  %K^-\pi^+\pi^-\pi^0$ and $K^+ K^-\pi^+\pi^-\eta$ Cross Sections Measured with
  %Initial-State Radiation,''
  Phys.\ Rev.\  D {\bf 76}, 092005 (2007)
  [arXiv:0708.2461 [hep-ex]].
  %%CITATION = PHRVA,D76,092005;%%


%\cite{Aubert:2006xw}
\bibitem{Aubert:2006xw}
  B.~Aubert {\it et al.}  [BABAR Collaboration],
  %``Observation of the exclusive reaction e+ e- --> Phi eta at s**(1/2) =
  %10.58-GeV,''
  Phys.\ Rev.\  D {\bf 74}, 111103 (2006)
  [arXiv:hep-ex/0611028].
  %%CITATION = PHRVA,D74,111103;%%

\bibitem{Adam:2004pr}
  N.~E.~Adam {\it et al.}  [CLEO Collaboration],
  %``Observation of 1- 0- final states from psi(2S) decays and e+ e-
  %annihilation,''
  Phys.\ Rev.\ Lett.\  {\bf 94}, 012005 (2005)
  [arXiv:hep-ex/0407028].
  %%CITATION = PRLTA,94,012005;%%

%\cite{Yao:2006px}
\bibitem{Yao:2006px} W.~M.~Yao \textit{et al.} {[}Particle Data
Group], %``Review of particle physics,''
 J.\ Phys.\ G \textbf{33}, 1 (2006). %%CITATION = JPHGB,G33,1;%%


%\cite{Feldmann:1999uf}
\bibitem{Feldmann:1999uf}
  T.~Feldmann,
  %``Quark structure of pseudoscalar mesons,''
  Int.\ J.\ Mod.\ Phys.\  A {\bf 15}, 159 (2000)
  [arXiv:hep-ph/9907491].
  %%CITATION = IMPAE,A15,159;%%



%\cite{Feldmann:1998vh}
\bibitem{Feldmann:1998vh} T.~Feldmann, P.~Kroll and B.~Stech,
%``Mixing and decay constants of pseudoscalar mesons,''
 Phys.\ Rev.\ D \textbf{58}, 114006 (1998) {[}arXiv:hep-ph/9802409],
%%CITATION = PHRVA,D58,114006;%%
 T.~Feldmann, P.~Kroll and B.~Stech, %``Mixing and decay constants of pseudoscalar mesons: The sequel,''
 Phys.\ Lett.\ B \textbf{449}, 339 (1999) {[}arXiv:hep-ph/9812269].
%%CITATION = PHLTA,B449,339;%%


%\cite{Donoghue:1986wv}
\bibitem{Donoghue:1986wv}
  J.~F.~Donoghue, B.~R.~Holstein and Y.~C.~R.~Lin,
  %``Chiral Loops In Pi0, Eta0 $\to$ Gamma Gamma And Eta Eta-Prime Mixing,''
  Phys.\ Rev.\ Lett.\  {\bf 55} (1985) 2766
  [Erratum-ibid.\  {\bf 61} (1988) 1527].
  %%CITATION = PRLTA,55,2766;%%

%\cite{Gilman:1987ax}
\bibitem{Gilman:1987ax}
  F.~J.~Gilman and R.~Kauffman,
  %``The Eta Eta-Prime Mixing Angle,''
  Phys.\ Rev.\  D {\bf 36}, 2761 (1987)
  [Erratum-ibid.\  D {\bf 37}, 3348 (1988)].
  %%CITATION = PHRVA,D36,2761;%%

%\cite{Akhoury:1987ed}
\bibitem{Akhoury:1987ed}
  R.~Akhoury and J.~M.~Frere,
  %``eta, eta-prime MIXING AND ANOMALIES,''
  Phys.\ Lett.\  B {\bf 220}, 258 (1989).
  %%CITATION = PHLTA,B220,258;%%

%\cite{Ball:1995zv}
\bibitem{Ball:1995zv}
  P.~Ball, J.~M.~Frere and M.~Tytgat,
  %``Phenomenological evidence for the gluon content of eta and eta-prime,''
  Phys.\ Lett.\  B {\bf 365}, 367 (1996)
  [arXiv:hep-ph/9508359].
  %%CITATION = PHLTA,B365,367;%%

%\cite{Klopot:2008ec}
\bibitem{Klopot:2008ec}
  Y.~N.~Klopot, A.~G.~Oganesian and O.~V.~Teryaev,
  %``Dispersive Approach to Abelian Axial Anomaly and $\eta-\eta'$ Mixing,''
  arXiv:0810.1217 [hep-ph].
  %%CITATION = ARXIV:0810.1217;%%







%\cite{Ball:1998sk}
\bibitem{Ball:1998sk} P.~Ball, V.~M.~Braun, Y.~Koike and K.~Tanaka,
%``Higher twist distribution amplitudes of vector mesons in {QCD}: Formalism
 %and twist three distributions,''
 Nucl.\ Phys.\ B \textbf{529}, 323 (1998) {[}arXiv:hep-ph/9802299].
%%CITATION = NUPHA,B529,323;%%


%\cite{Kroll:2002nt}
\bibitem{Kroll:2002nt} P.~Kroll and K.~Passek-Kumericki, %``The two-gluon components of the eta and eta' mesons to leading-twist
 %accuracy,''
 Phys.\ Rev.\ D \textbf{67}, 054017 (2003) {[}arXiv:hep-ph/0210045].
%%CITATION = PHRVA,D67,054017;%%

%\cite{Gorsky:1985aw}
\bibitem{Gorsky:1985aw} A.~S.~Gorsky, %``Form-Factor Pi Rho Gamma In Perturbative QCD,''
%\href{http://www.slac.stanford.edu/spires/find/hep/www?irn=1445979}{SPIRES entry}
\textit{Moscow Inst. Theor. Exp. Phys. Gkae - ITEF-85-071. }

\end{thebibliography}
\end{document}